\def\aj{{AJ}}

\def\apj{{ApJ}}

\def\mnras{{MNRAS}}

\documentclass[cup5b]{caps}
\usepackage{graphicx}
\usepackage{amssymb}
\usepackage{ociwsymp1e}
\HeadText{D. J. E. Floyd}

\begin{document}

\pagenumbering{arabic}

\author[]{D. J. E. FLOYD \\ Royal Observatory Edinburgh}
\chapter{Luminous Quasars and Their Hosts:\\Accretion at the Limit?}
\begin{abstract}
  We present the results of our recent
  Hubble Space Telescope imaging study, in which we have successfully
  deconvolved host and nuclear flux for the some of the most luminous
  quasars in the Universe.
  Host morphologies have been recovered for each of our 17-strong
  sample. From these fits, we have estimated Black Hole masses through
  extrapolation of the Black Hole - Spheroid mass relation and begun to
  address the complicated issue of fuelling vs black hole mass in
  determining quasar luminosity:
  We find that the order-of-magnitude 
  increase in luminosity is due to
  increased black hole size, and to increased fuelling efficiency, in
  roughly equal measure. The brightest objects are found to radiate at
  or near the Eddington limit.
\end{abstract}

\section{Introduction}
\label{sec-intro}
As the most powerful class of AGN, quasars
provide a striking example of the complex interelationship between
galaxies and the supermassive black holes in their centres. 
Thanks largely to the Hubble Space Telescope (HST), the last five years
have seen huge advances in our understanding of the host galaxies
of the nearest ($z<0.3$) quasars. 

\begin{figure}
\includegraphics[width=\columnwidth]{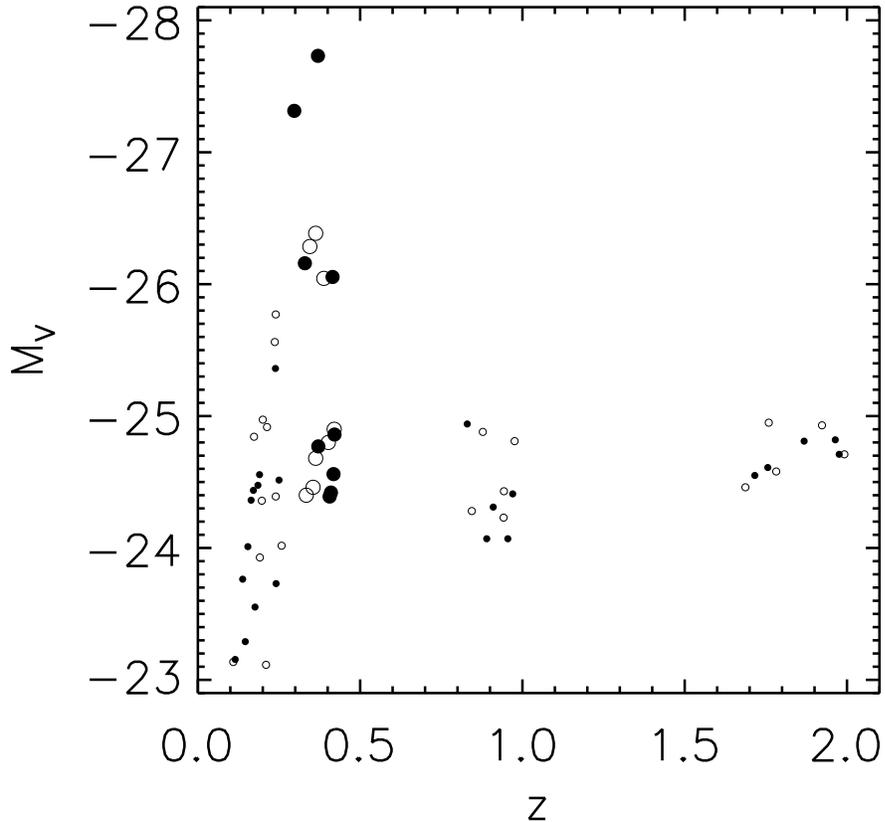}
\label{fig-samp}
\caption{Absolute magnitude versus redshift for quasars observed to
date in our HST host-galaxy imaging programmes. Filled circles
represent radio-quiet quasars (RQQs) and open circles radio-loud
quasars (RLQs). Our earlier work (small symbols) concentrated mainly
on quasars of relatively low luminosity (typically $M_{V}>-25$) in
three redshift regimes ($z \simeq 0.2, 1$ \& 2), allowing us to probe
the evolution of the host galaxies over a large fraction of cosmic
history (Mclure et al. 1999; Kukula et al. 2001; Dunlop et
al. 2002). The current study (large symbols) is designed to explore an
orthogonal direction in the $M_{V} - z$ plane, by sampling a large
range of quasar luminosities at a single redshift, $z\simeq 0.4$. This
is the lowest redshift at which very luminous quasars (those with
$M_{V} < -27$) can be found, comparable to the most luminous quasars
in the high-redshift universe.}
\end{figure}

Most quasar host studies have hitherto concentrated on quasars of
relatively low luminosity, because it is much easier to
disentangle the contributions of host and active nucleus. 
However, the quasar population includes objects as bright as
$M_{V}\sim -30$, and the majority of currently known high-$z$ quasars
belong to the high end of the luminosity function (largely due to the
degeneracy inherent in any flux-limited sample).
The aim of the current study is to break this degeneracy between quasar
luminosity and redshift by studying a sample of quasars at a single
redshift, but spanning an appreciable fraction of the quasar
luminosity range (fig.~\ref{fig-samp}). 
The lowest redshift at which this can be done is $z\sim0.4$.

Not only will this allow us to explore the relation between quasar
luminosity and the properties of their host galaxies, but the most
luminous objects in this programme will also provide a low redshift
baseline against which to compare the hosts of luminous high-$z$
quasars in future studies.

\section{Observational Strategy}
\label{sec-obstrat}
Imaging of quasars must be carefully choreographed.
We require extremely high dynamic range in order to accurately
characterise both the low-surface-brightness features of the host, and
the critical, highly luminous core region.
HST's stable pointing accuracy allows us to take
several deep integrations, and splice shorter ones directly into the
core where saturation has occurred.

A thorough knowledge of the PSF is essential, and across an immense
dynamic range. To this end, we combine the sub-pixel centred
models of TINYTIM (Krist, 1999) with extremely deep stellar PSF's to
accurately model all of the scattered flux in the wings.
(see Floyd et al. (2002) for full details).

We minimise the nuclear to host ratio, through careful filter
selection. We observe in the 
$V$-band in the quasar's restframe (longwards of
the 4000\AA~break), and avoid any strong emission lines from the
circumnuclear region (H$\alpha \lambda 6563$; [O{\sc iii}]$\lambda
5007$).
By excluding such emission from the images, we obtain a cleaner
picture of the distribution of starlight in the hosts.

\section{Image Analysis \& Modelling}
\label{sec-model}
We then fit using the model of Sersic (1968):
\begin{equation}
\label{eqn-sersic}
\mu(r)=\mu_{0} \exp \left \{ -\left( \frac{r}{r_{0}} \right) ^{\beta}
\right \} 
\end{equation}
where $\mu(r)$ describes an azimuthally-symmetric distribution, which can
be projected onto a generalised elliptical coordinate system to allow
for different eccentricities and orientations of the host galaxy.
Additional flux is added to the central pixel to model the unresolved
nuclear component.

We therefore have a 6-dimensional model:
\begin{itemize}
\item $L_{n}=$ nuclear luminosity
\item $\mu_{1/2}=$ host surface brightness
\item $R_{1/2}=$ host half-light radius
\item $\Theta=$ position angle
\item $\xi=\frac{a}{b}=$ axial ratio
\item $\beta-$ parameterising host profile shape
\end{itemize}
The model is convolved with the PSF and compared to the real quasar
image.  We choose weighting on an individual pixel basis,
assigning each pixel an error based on its Poisson 
noise. We use the downhill simplex method as a robust technique for
$\chi^{2}$ minimisation. 

\section{Results}
\label{sec-res}
We began by fixing the morphology {\em a priori} to either a bulge
($\beta=\frac{1}{4}$) or a disc ($\beta=1$). The best fit bulge and
disc models are then compared, and the full 6-dimensional model is
run. 

Overall, the morphologies of the hosts in the current sample agree
well with the results of Dunlop et al. (2002). 
We find a clear preference for
elliptical hosts in all cases bar three (three out of
the five least luminous RQQs - in each case deconvolved nuclear luminosity
is close to the dividing line with Seyfert nuclei).
For the majority of objects the morphological preference was confirmed
by the variable-$\beta$ model. The hosts are also found to follow a
Kormendy relation similar to that displayed by nearby quiescent
ellipticals (fig.~\ref{fig-korm}).

\begin{figure}
\includegraphics[width=\columnwidth]{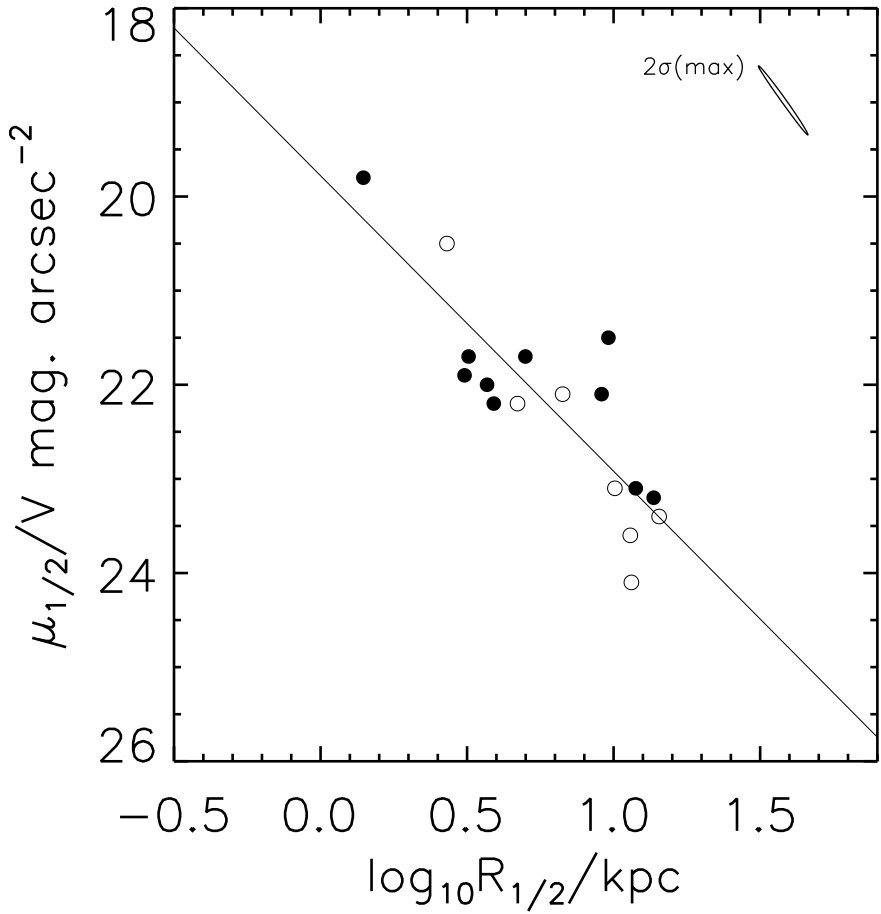}
  \caption{The half-light radius ($R_{1/2}$) vs surface brightness
  ($\mu_{1/2}$) at the half-light radius for our sample. Filled
  circles represent RQQ hosts, open circles RLQ hosts. The solid line
  shows the best fit Kormendy relation to the sample and has the form
  $\mu_{1/2}=19.8+3.14log_{10}R_{1/2}$. The narrow ellipse in the top
  right corner of the plot shows the $2\sigma$ error contours for the
  least robust model fit in the current sample.}
\end{figure}

\section{Black Hole Mass}
\label{sec-BH}
It has become increasingly clear from studies of
inactive galaxies that black hole and galaxy formation and growth are
intimately linked processes
(Magorrian et al. 1998, Gebhardt et al. 2000, Merritt \& Ferrarese
2001). We use luminosities from our galaxy models to
estimate host masses, (assuming mass-to-light
ratio of an early-type galaxy).
We then use the Magorrian Relation to estimate Black Hole mass. 
From the black hole mass we can
then calculate the theoretical Eddington luminosity of each object,
and compare this with the actual luminosity of
the quasar nucleus obtained from our model fit. The results of this
work can be compared directly with those from other recent host-galaxy
studies (fig.~\ref{fig-nuchost}). We find that all objects are
radiating at or below the Eddington limit.

Several ground based studies have resulted in points that appear to
the right of the Eddington limit (e.g. Percival et al. 2001),
but we believe that this is due largely to poor seeing
preventing accurate disentanglement of host and nuclear
flux. We obtained HST archive images of 3 objects from the sample of
Percival et al. (2001), and analysed them as for our sample (Floyd et
al. 2002). We also included the 2 cases for which Percival et
al. unambiguously uncover an Elliptical host. These objects
are included in fig.~\ref{fig-nuchost}, marked by stars. This
substitution of ground based data results in a remarkable agreement
with Eddington limited accretion.

\begin{figure}
\includegraphics[width=\columnwidth]{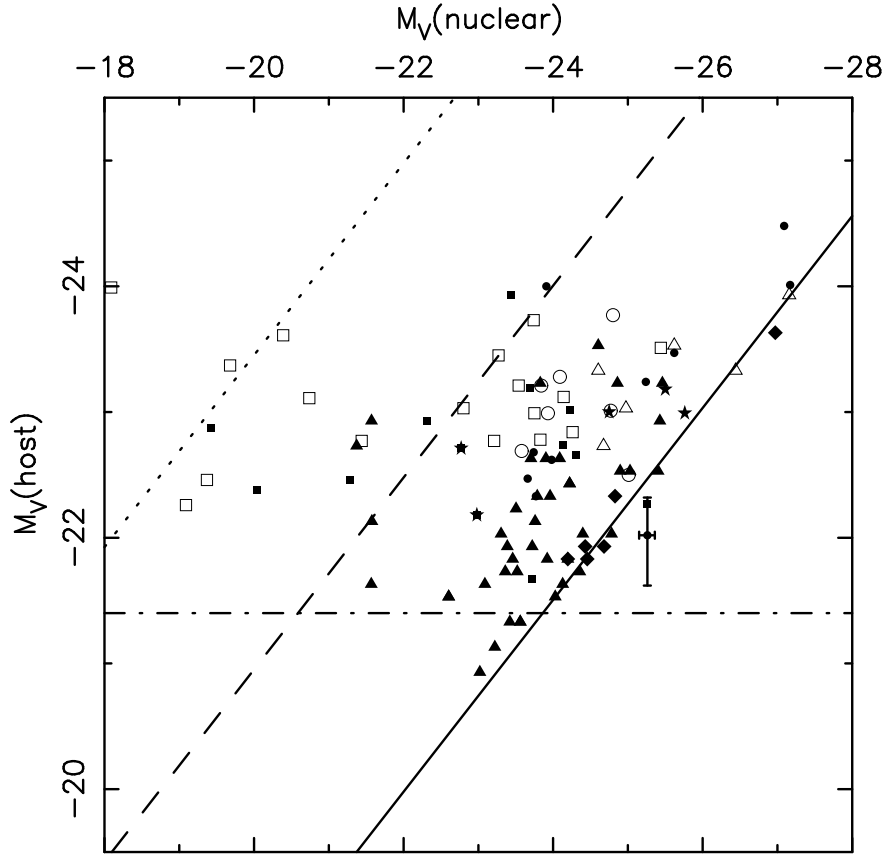}
  \caption{Host vs Nuclear magnitudes for quasars from this work
  (circles), along with those from Dunlop et al. (2002) - squares;
  McLeod \& Rieke (1994) - triangles; McLeod \& McLeod (2001) - diamonds;
  and also 5 objects from Percival et al. (2001) - stars. The solid line
  illustrates the Eddington limited accretion rate, with the dashed
  and dotted lines denoting the 10\% and 1\% levels respectively. The
  one object from our sample that appears to be accreting at somewhat
  higher than the Eddington limit is the luminous quasar 1252$+$020,
  and is the one object for which we do not believe that we have
  successfully disentangled host and nuclear flux (see Floyd et
  al. (2002) for more details).} 
\end{figure}

\section{Conclusions}
\label{sec-conc}
\begin{itemize}
\item Quasar hosts are large, luminous ellipticals ($L > L^{\star}$),
  consistent with radio galaxies; we are well into
  the exponential tail of the galaxy luminosity function.

\item The factor of 10 increase observed in quasar nuclear luminosity
  appears to be due to a combination of increasing fuelling efficiency
  and increasing black hole mass. 

\item The Eddington limit seems to impose the maximum luminosity for
  an AGN in a given host. An Elliptical like M87, with
  $M_{V}\approx-25$, would therefore be capable of hosting an AGN as
  luminous as $M_{V}\approx-29$. 

\item Observations to yield direct measurements of black hole
  mass (e.g. H$\beta$ linewidths) form a natural next step. These will
  allow more accurate determination of black hole mass and accretion
  efficiency. 

\end{itemize}

\section{The Future}
\label{sec-fut}
It is clear now that both engine size and fuelling efficiency are key
factors in determining a quasars optical luminosity. However, their
contributions are still unclear. We now need tests of
the Spheroid - Black Hole Mass relation for the high end of the
galaxy luminosity function. More direct measures of Black Hole Mass
(e.g. from BLR line widths) will assist. The 2D modelling technique
also offers us an automated morphological tool for the study of
inactive galaxies, as well as a wide range of active ones.


\begin{thereferences}{}

\bibitem{dunlop+02}
Dunlop, J.S. et al. 2002, \mnras, in press astro-ph/0108397 

\bibitem{floyd+02}
Floyd, D.J.E. et al. 2002, \mnras, in preparation.

\bibitem{gebhardt+00}
Gebhardt, K. et al. 1998, \apj, 539, L13

\bibitem{tinytim}
Krist, J., TinyTim User Manual, 1999

\bibitem{kukula+01}
Kukula, M.J., et al. 2001, \mnras, 326, 1533

\bibitem{mcleodrieke94}
McLeod, K.K. \& Rieke, G.H. 1994, \apj, 431, 137

\bibitem{mcleod+99}
McLeod, K.K., Rieke, G.H. \& Storrie-Lombardi, L.J., 1999, \apj, 511,
L67 

\bibitem{mcleod01}
McLeod, K.K. \& McLeod, B.A., 2001, \apj, 546, 782

\bibitem{mclure+99}
McLure, R.J., et al. 1999, \mnras, 308, 377

\bibitem{magorrian+98}
Magorrian, J. et al. 1998, \aj, 115, 2285

\bibitem{merrittferrarese01}
Merritt, D. \& Ferrarese, L. 2001, \mnras, 320, L30

\bibitem{percival+01}
Percival, W.J., Miller, L., McLure, R.J. \& Dunlop, J.S. 2001, \mnras,
322, 843

\bibitem{sersic68}
Sersic, J.L. 1968, Atlas de Galaxes Australes (Cordoba: Observatorio
de Universidad de Cordoba).  

\end{thereferences}
\end{document}